\documentclass[3p,preprint,twocolumn]{elsarticle}

\usepackage{lineno,hyperref}
\usepackage{bm}
\usepackage{placeins}
\usepackage{array}
\usepackage{caption}
\usepackage{cancel}

\usepackage{booktabs,makecell,multirow,array,colortbl,dcolumn,stfloats} 

\usepackage{framed} 
\usepackage{multicol} 
\usepackage{nomencl} 
\makenomenclature
\renewcommand*\nompreamble{\begin{multicols}{2}}
\renewcommand*\nompostamble{\end{multicols}}

\modulolinenumbers[5]

\usepackage{color}
\usepackage[size=footnotesize]{todonotes} 
\definecolor{rot}{RGB}{194,0,0}
\definecolor{gruen}{RGB}{0,146,0}
\usepackage{relsize}
\usepackage[utf8]{inputenc}
\usepackage{mathtools}
\usepackage{amsmath}
\usepackage{amssymb}
\newcommand{\ve}[1]{\boldsymbol{#1}} 
\newcommand{\te}[1]{\boldsymbol #1} 

\newcommand{\inte}[3]{\mathop{\int}_{ #1} #2 \; \mathrm{d} #3} 

\newcommand{\tr}{\mathrm{tr}\,}

\newcommand{\calW}{{\cal W}}
\newcommand{\calB}{{\cal B}}

\newcommand{\Gc}{G_\mathrm{c}}
\newcommand{\sig}{\te{\sigma}}

\newcommand{\eps}{\te{\varepsilon}}
\newcommand{\epse}{\te{\varepsilon}^\mathrm{e}}

\newcommand{\psie}{\psi^\mathrm{e}}

\journal{International Journal of Fatigue}

\begin{document}

\begin{frontmatter}

\title{Fatigue crack growth in anisotropic aluminium sheets -- phase-field modelling and experimental validation}


\author[mymainaddress]{Martha Kalina}

\author[mysecondaddress]{Vanessa Schöne}

\author[mymainaddress]{Boris Spak}

\author[mysecondaddress]{Florian Paysan}

\author[mysecondaddress]{Eric Breitbarth}

\author[mymainaddress,mythirdaddress]{Markus K\"{a}stner\corref{mycorrespondingauthor}}
\cortext[mycorrespondingauthor]{Corresponding author}
\ead{markus.kaestner@tu-dresden.de}

\address[mymainaddress]{Chair of Computational and Experimental Solid Mechanics, TU Dresden, Dresden, Germany}
\address[mysecondaddress]{Institute of Materials Research, German Aerospace Center (DLR), Köln, Germany} 
\address[mythirdaddress]{Dresden Center for Computational Materials Science (DCMS), TU Dresden, Dresden, Germany}

\begin{abstract}
Fatigue crack growth is decisive for the design of thin-walled structures such as fuselage shells of air planes. The cold rolling process, used to produce the aluminium sheets this structure is made of, leads to anisotropic mechanical properties. In this contribution, we simulate the fatigue crack growth with a phase-field model due to its superior ability to model arbitrary crack paths. A fatigue variable based on the Local Strain Approach describes the progressive weakening of the crack resistance. Anisotropy regarding the fracture toughness is included through a structural tensor in the crack surface density. The model is parameterised for an aluminium AA2024-T351 sheet material. Validation with a set of experiments shows that the fitted model can reproduce key characteristics of a growing fatigue crack, including crack path direction and growth rate, considering the rolling direction.
\end{abstract}

\begin{keyword}
Phase-field \sep Fracture \sep Fatigue crack growth \sep Aluminium sheets  \sep Anisotropy
\end{keyword}

\end{frontmatter}

\section{Introduction}

Understanding the fatigue crack behaviour of materials is essential for assessing the service life and damage tolerance of thin-walled structures such as aircraft fuselages. The so-called damage tolerance concept for ensuring structural integrity relies on the ability of a defective structure to withstand the expected loads until this defect can be detected and repaired by scheduled inspection or non-safety-relevant failure \cite{tavares_overview_2017}. To evaluate the service life, Wöhler curves provide information on crack initiation and damage accumulation \cite{schijve_fatigue_2003}. Then, the range of stable fatigue crack growth (FCG) is determined by crack propagation tests under consideration of small-scale yielding and relates the crack tip stress to the crack growth rate \cite{astme647-15_standard_2015}. These experimentally determined crack propagation curves serve as basis for fitting FCG describing models, for example FASTRAN \cite{newman_fastran2_1992} or NASGRO \cite{_nasgro_2010}, to predict the fatigue life and fatigue crack behavior of structures under variable loads \cite{sagrillo_elastic_2022}. The unified two-parameter driving force model for fatigue crack growth analysis \cite{Noroozi2005} establishes a correlation between the maximum stress intensity factor, the stress intensity range and the elasto-plastic stress-strain field of the crack tip. Therefore, the UniGrow model can also account for overload effects or residual stresses from cyclic plasticity near the crack tip \cite{Mikheevskiy2015}. Additional modifications extend the UniGrow model to the range of short fatigue crack growth \cite{Bang2019}.

The fatigue crack propagation behavior of ductile materials is controlled by an interaction between damaging mechanisms in front of the crack tip and extrinsic shielding mechanisms due to the plastic wake \cite{american1971damage}. Damaging mechanism include, for example, intergranular fracture or micro-void nucleation in front of the crack tip, which are material-specific and are influenced by the crack driving force as well as the crack tip stress field \cite{needleman_analysis_1987}. Shielding mechanisms such as crack deflection or crack closure reduce the crack tip loadings and thus have a retarding effect \cite{ritchie_no_1999}. 

\begin{figure} [h] 
    \centering
	\def\svgwidth{\linewidth}\small{
		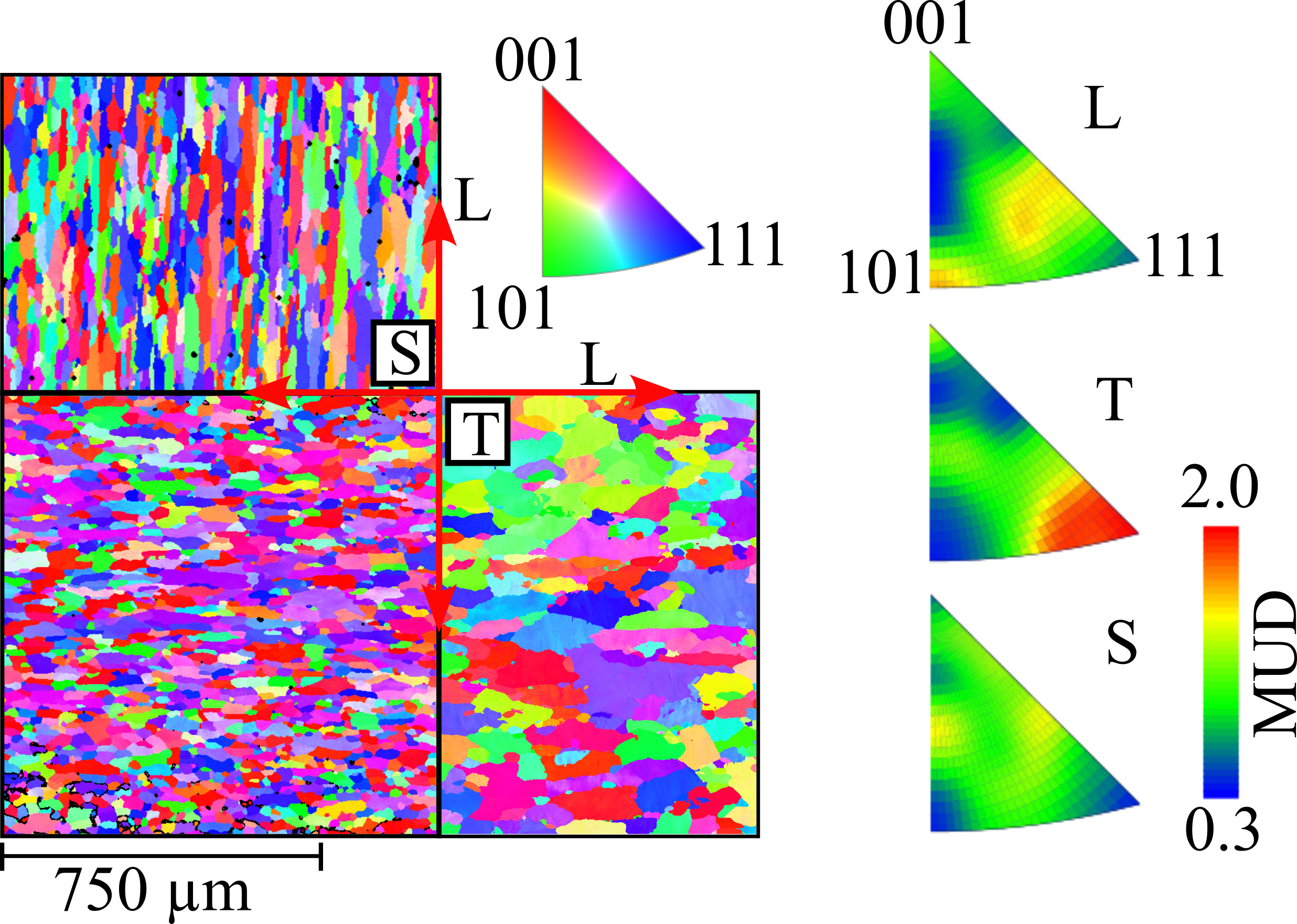}
	\caption{{EBSD scan of a rolled 2\,mm thick AA2024-T351 sheet. Grain elongation along the rolling direction L is visible. Grain length to height ratio for LT/TS/LS:  2.1/2/3.3.}}
	\label{fig:microstruct}
\end{figure}

The aluminum alloy AA2024-{T351} studied here is widely used for airplane fuselages, which can be explained by its excellent damage tolerance behaviour \cite{dursun_recent_2014}. During the manufacturing process, the base cast material is rolled along the longitudinal axis. Material properties show a rolling direction-dependent anisotropy due to grain elongation and an orientation of the texture \cite{wei_influence_2014,xia_texture_2018}, as can be seen in the EBSD scan of AA2024-{T351} in Fig.~\ref{fig:microstruct}. As a consequence, many investigators reported anisotropic strength properties 
\cite{gilmour_influence_2004,barlat_sixcomponent_1991,seidt_plastic_2013}. 
Regarding the static crack resistance, Amstutz et al. \cite{amstutz_effects_1997} tested AA2024-{T351} sheet material with a thickness of 2.3\,mm in L-T (crack perpendicular to grain elongation) and T-L (crack parallel to grain elongation) orientation under Mode-I loading. They pointed out that the resulting fracture surface depends on the orientation. Specimen in L-T orientation showed a higher crack resistance compared to the T-L orientation. This is in agreement with tests by Johnston and Newman \cite{johnston_fracture_2001} on 1.6\,mm thin AA2024-{T351} sheets. 

Regarding FCG behavior, Bergner \cite{bergner_new_2000} performed constant-amplitude Mode-I fatigue tests on MT160 specimen with a thickness of 1.6\,mm, made from AA2024 alloys. His tests revealed a dependency of the crack propagation behavior, not only on the temper condition, but also on the orientation, showing an increased crack propagation rate for cracks in T-L orientation. This was confirmed by Da Cunha and de Lima \cite{carvalhodacunha_influence_2017}:
A slightly extended lifetime can be seen for L-T oriented specimens. This also applies for crack growth behavior under variable test amplitudes representing sequences of aircraft flight load spectra as the test setup was given by Kocańda and Jasztaal \cite{kocanda_probabilistic_2012}. They tested 3\,mm thick Middle Tension (MT) test sheets.
L-T-oriented specimens endure a higher number of load sequences until failure. 

Traditional methods to predict and simulate FCG such as the NASGRO equations, cohesive zone models \cite{kuna_numerische_2010} and XFEM \cite{moes_finite_1999} involve certain drawbacks. Usually, the crack path has to be known a priori or is intricately described with additional elements or enriched ansatz functions. The phase-field method, though, is a powerful framework to model crack initiation, propagation and arbitrary crack paths in a straight-forward way. 
This is particularly important for anisotropic materials under realistic loads where the crack path cannot be previously known or predicted easily.
The crack geometry is described with an additional field variable -- the phase-field variable -- in a regularised manner, forming a coupled problem. See \cite{ambati_phasefield_2015} for an overview of the original phase-field models for brittle static fracture. Later, the method was extended to ductile \cite{alessi_comparison_2018} and rate-dependent \cite{dammass_unified_2021} material behaviour. A comparison of different approaches to model anisotropic \textit{static} fracture is given in \cite{scherer_assessment_2022}. 

More recently, the method has also been applied to fatigue fracture, see \cite{kalina_overview_2023} for a comparison of various phase-field fatigue models. They include brittle and ductile material behaviour, high and low cycle fatigue as well as various material classes such as metals, elastomers and concrete. However, to the best of the authors knowledge, no phase-field fatigue model for anisotropic material behaviour has been published yet. In order to model the direction-dependent fracture behaviour of rolled aluminium sheet material, we introduce an extension to the phase-field fatigue model by Seiler et al. \cite{seiler_efficient_2020} to anisotropic fracture toughness, based on the anisotropic static phase-field formulation {by} Teichtmeister et al. \cite{teichtmeister_phase_2017}. {Thereby, we establish a versatile framework allowing to simulate fatigue crack initiation and growth independent of the structure's geometry under arbitrary loading conditions, including crack deflection and change in the crack growth rate due to direction-dependent material qualities. The model is here parameterised for an aluminium material, yet it can be applied to any sheet metal material.}

This paper is structured as follows: Chapter 2 presents the FCG experiments on MT specimen which show a dependency on the rolling direction. Chapter 3 introduces the phase-field model for anisotropic fatigue fracture to simulate these phenomena. The experimental determination of model parameters -- including elastic, cyclic and fracture parameters -- is presented in Chapter 4. Finally, the phase-field model is parameterised in Chapter 5 and subsequently used for numerical studies of the fracture phenomena previously observed in the MT experiments. The paper ends with a conclusion.

\section{Fatigue fracture in aluminium sheets}
\label{sec:fatfrac}

\begin{figure} [h] 
    \centering
	\def\svgwidth{\linewidth}\small{
		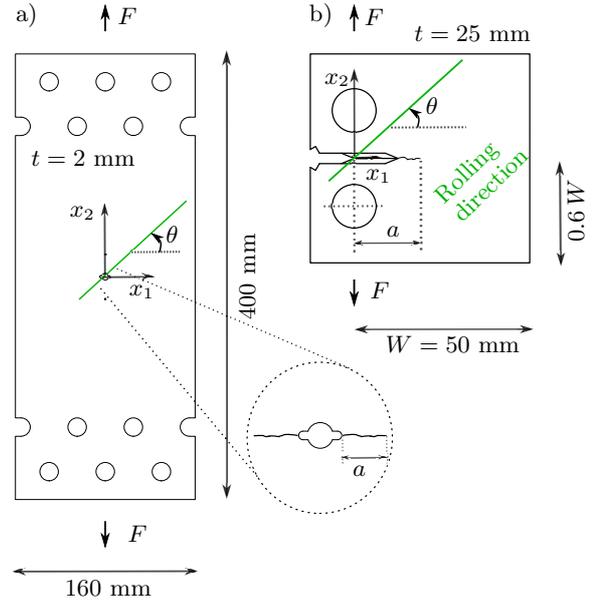}
	\caption{\textbf{(a)} MT specimen. \textbf{(b)} CT specimen. }
	\label{fig:specimen}
\end{figure}

{
\subsection{Experimental setup}

An aluminum alloy A2024-{T351} (AlCu4Mg1) was used. Rolled sheets with a thickness of 2\,mm were procured by a commercial supplier and test specimens were machined.} The fatigue crack behavior was determined using MT160 specimens, see Fig.~\ref{fig:specimen}a, in three orientations: 
\begin{itemize}
    \item $\theta=0$°: T-L orientation, crack growth parallel to grain elongation (Load in T-, FCG in L-direction)
    \item $\theta=90$°: L-T orientation, crack growth perpendicular to grain elongation (Load in L-, FCG in T-direction)
    \item $\theta=45$°: crack growth 45° to grain elongation 
\end{itemize}
The crack propagation tests were carried out according to ASTM E674-15 \cite{astme647-15_standard_2015}. The specimen is loaded at constant force amplitude with $F_{\max} =15$\,kN and load ratios $R =$ \mbox{$ F_{\min}/F_{\max} =\{0.1, 0.3, 0.5\}$} on a uniaxial servo-hydraulic testing machine.
Using a sinusoidal load with constant amplitude and a test frequency of 20\,Hz, fatigue crack propagation was initiated from a centric initial notch with a length of $2a = 16$\,mm, {sawed symmetrically into the specimen}. The potential difference  was measured between two pins located 10\,mm below and above the fatigue crack \mbox{($I_\mathrm{DC-POT}$ = 100\,A = const.,} \mbox{$U_\mathrm{DC-POT}$ = 60\,mV)} to calculate the crack length by means of the Johnson equation \cite{johnson_calibrating_1965}. To determine the crack growth rate, the secant method was used. The range of stress intensity factors $\Delta K$ were calculated according to ASTM E647-15 and the formula provided by Feddersen \cite{feddersen_discussion_1966}, leading to fatigue crack propagation curves ($\mathrm{d}a/\mathrm{d}N-\Delta K$) shown in Fig.~\ref{fig:exp_results}.

Additionally, full-field displacement and strain information at the surface were captured using a GOM Aramis 12M 3D digital image correlation (DIC) system. Using a black and white stochastic pattern and a facet size of 25$\times$25 pixel with a facet distance of 16\,pixels (1\,pixel = 0.045\,mm), it is possible to achieve a spatial resolution of 0.72\,mm. Images were taken each $\Delta a = 0.5$\,mm crack propagation for maximum, mean and minimum load of the cycle. For calculating the displacements, we use the GOM Aramis Professional V2020 software. Based on this displacement field, a trained artificial neural network is used to determine the crack tip position \cite{melching_explainable_2022,strohmann_automatic_2021} and the crack path. 

\subsection{Fatigue crack propagation data}

\begin{figure} [p] 
	\def\svgwidth{\linewidth}\small{
		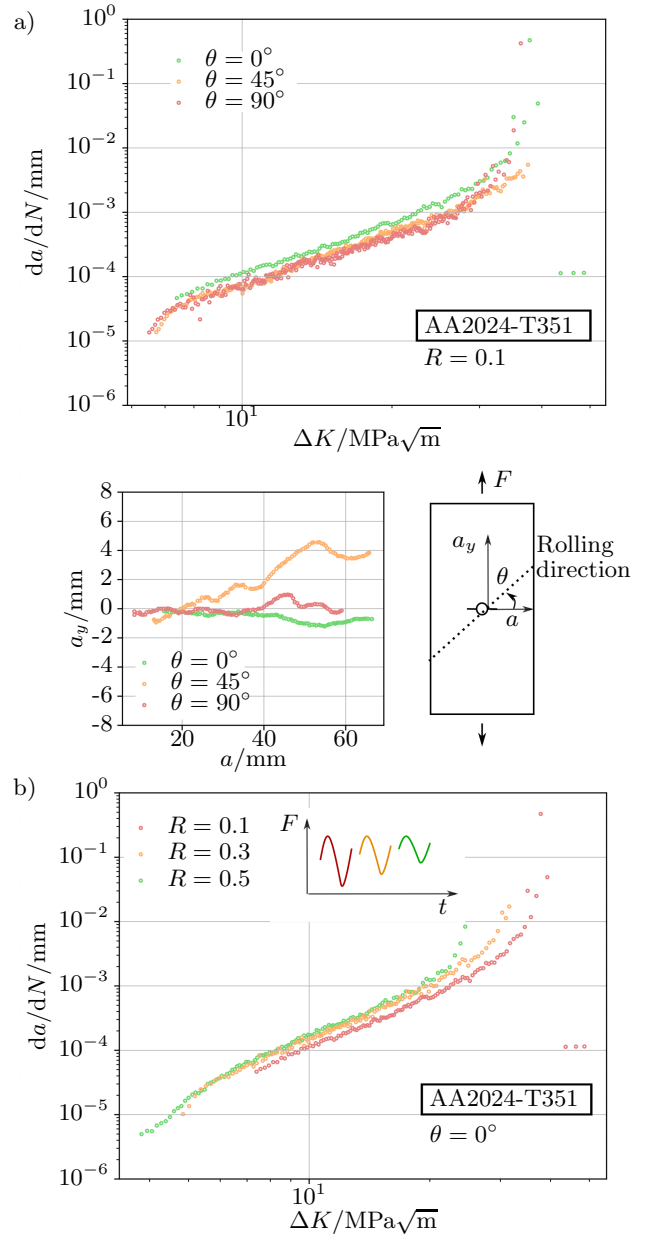}
	\caption{Experimental results of FCG experiments with MT specimens. \textbf{(a)} Variation of the angle between crack and sheet rolling direction $\theta$. The fatigue crack propagation rate (upper) varies little. The crack path (lower) inclines for $\theta=45^\circ$. \textbf{(b)} Variation of the load ratio $R$ between load minimum and maximum of a load cycle for $\theta=0$°. Increased $R$-ratio, i.\,e. increasing load minimum while keeping maximum load constant, reduces the fatigue crack propagation rate. $a$ is the crack length in $x_1$-direction, $a_y$ the crack position in $x_2$-direction.  }
	\label{fig:exp_results}
\end{figure}

The $\mathrm{d}a/\mathrm{d}N-\Delta K$ FCG  curves for $R=0.1$ are shown in Fig.~\ref{fig:exp_results}a. The highest crack propagation rate can be found for $\theta=0$° (T-L) which corresponds to crack growth parallel to grain elongation. The crack propagation rates for $\theta=45$° and $\theta=90$° (L-T) are { almost} similar. When comparing the resulting crack path for the right side of the cracked MT specimen, it becomes apparent that the resulting path depends on the orientation, though. $\theta=0$° shows a predominant straight path with growing crack length, whereas the crack path of $\theta=90$° is characterized by a zig-zag path $\pm1$\,mm around the initial crack path. The path of $\theta=45$° reveals the highest crack path slope where it is clearly visible that the crack path is strongly influenced  by the rolling direction. 

FCG rates are lower in L-T direction since there are more grain boundaries to cross perpendicular to the elongated grains, blocking dislocations \cite{jesus_influence_2022}. Instead, cracks often tend to follow grain boundaries since these are weaker areas due to the precipitation of the supersaturated mixed crystal and the collision of several crystals, leading to crystallographic defects. Furthermore, the rolling process changes the texture, making the crystallites align in rolling direction. The oriented slip systems favour cracks parallel to the rolling direction \cite{wei_influence_2014,xia_texture_2018}.

Regarding the Paris curves for different R-ratios $R=\{0.1, 0.3, 0.5\}$ and $\theta=0$° in Fig.~\ref{fig:exp_results}b, { the crack propagation rates are also very similar when being plotted over the range of the stress intensity factor $\Delta K$, with $\theta=0$° showing the highest crack propagation rate. }

In the following, these results are to be reproduced with phase-field simulations.

\section{Phase-field model for fatigue crack propagation in anisotropic material}

The fatigue crack propagation experiments with anisotropic aluminium material shall be simulated with a phase-field model. In order to do so, an existing phase-field model for fatigue fracture by {Seiler} et al. \cite{seiler_efficient_2020} is extended to cover anisotropic material behaviour. In the following, the original model is recapitulated first, the extension to anisotropy is presented thereafter.

\subsection{Isotropic material behaviour}

The underlying isotropic phase-field fatigue model \cite{seiler_efficient_2020} is formulated in a small strain setting with linear elastic material behaviour. The considered domain in $\mathcal{B}\in\mathbb{R}^n$ with a boundary $\partial\mathcal{B}$ and material points defined by their location $\ve{x}$ at time $t$. The strain $\eps(\ve{x},t)$ is defined by $\eps=1/2\,(\nabla\ve{u} + \nabla\ve{u}^\top)$ with  displacement $\ve{u}(\ve{x},t)$. Fractured material can be described in a regularised manner with the phase-field variable $d(\ve{x},t)$, thereby $d=0$ indicates intact material and $d=1$ fully broken material. {The regularisation width is controlled by the length scale parameter $\ell$, chosen as a compromise between a narrow crack profile approximating the ideal sharp crack ($\ell\rightarrow0$) and a preferably coarse mesh.} The cyclic loading and damage history of the material point is described by the accumulating fatigue variable $D\in[0,1]$.

\subsubsection*{Energy functional}

The total energy density 
\begin{equation}
    W(\eps,d,\nabla d;D)  = g(d)\psie(\eps) + \alpha(D)\,\Gc\gamma_\mathrm{iso}(d,\nabla d)
\end{equation}
consists of an elastic part with the free energy density 
\begin{equation}
    \psie = \frac{1}{2}\lambda \,\tr^2(\eps) + \mu\,\tr\left(\eps\right)
\end{equation}
degraded by the degradation function \mbox{$g(d)=(1-d)^2$}
\footnote{While a split of the elastic energy density into a degraded tensile and an undegraded compressive part can be included in a straightforward manner, it is not applied in this case due to the numerical examples not requiring it.}
and the regularised fracture surface density
\begin{equation}
    \gamma_\mathrm{iso}(d,\nabla d) = \frac{d^2}{2\ell} + \frac{\ell}{2}|\nabla d|^2.
\end{equation}
The fracture toughness $\Gc$ is the parameter characterising the materials resistance against cracks.
The fracture energy is degraded by the fatigue degradation function $\alpha\in[\alpha_0,1]$ being
\begin{equation}
    \alpha(D) = (1-\alpha_0)(1-D)^\xi + \alpha_0
\end{equation}
with parameters $\alpha_0$ and $\xi$, describing the ongoing weakening of the material due to cyclic stressing in form of progressively reducing the fracture toughness.

The stress is defined as 
$\sig =  g(d) \, \partial\psie(\eps)/\partial\eps$. Given the volume force $\bar{f}$ and the traction vector $\bar{\tau}$, the total energy of the body is
\begin{align} \label{eq:Pi}
    \Pi(\eps,d,\nabla d;D) = & \inte{\calB}{W(\eps,d,\nabla d;D)}{v} \nonumber \\
    & - \inte{\calB}{\bar{f}\cdot\ve{u}}{v} - \inte{\partial\calB}{\bar{\tau}\cdot\ve{u}}{a}.
\end{align}

\subsubsection*{Derivation of model equations through variational principle}

The model equations of the coupled problem can be derived by the means of the variational principle
\begin{equation} \label{eq:varprin}
    \{ \ve{u}, d \} = \arg \left\{ \min_{\ve{u}\in \calW_{\bar{u}}} \min_{\ve{d}\in \calW_{d}} \Pi(\eps,d,\nabla d;D) \right\}
\end{equation}
with conditions 
\begin{align}
    \calW_{\bar{u}} &= \left\{ \ve{u}\in \mathbb{R}^3 \, | \, \ve{u} =  \bar{\ve{u}} \text{ on } \partial\calB^\mathrm{D} \right\} \\
    \calW_d &= \{ d\in\mathbb{R} \, | \, \dot{d}\geq0 \}
\end{align}
for prescribed displacements $\bar{\ve{u}}$ on {Dirichlet} boundaries $\partial\mathcal{B}$ and the irreversibility condition for the phase-field. The variational derivative $\delta_u\Pi$ with respect to $\ve{u}$ eventually yields the balance of momentum
\begin{equation}
    \nabla\cdot\sig + \bar{\ve{f}} = \ve{0} \text{ in } \calB 
\end{equation}
with the boundary condition $ \sig\cdot\ve{n}=\ve{\bar{t}}$ on $\partial\calB^\mathrm{N}$.
Likewise, $\delta_d\Pi$ yields the weak form of the phase-field problem
\begin{align} \label{eq:weakd}
\delta_d \Pi = &\int_{\calB}\left\{\left[g'(d)\psie  + \alpha(D) \frac{\Gc}{\ell}d  
\right] \delta d\, \right. \nonumber  \\
& \quad + \alpha(D)\Gc\ell \,\nabla d \cdot \delta \nabla d \Big\}\,\mathrm{d}v = 0, 
\end{align}
further demanding $\dot{d}\geq0$. Its local form is the phase-field evolution equation
\begin{align} \label{eq:ev}
2(1-d)\max_{\tau\in[0,t]}\psie(\tau) = & \, \Gc \alpha(D)\left(\ell\Delta d - \frac{d}{\ell} \right) \nonumber \\
& + \Gc\ell\nabla d\cdot\nabla \alpha(D)   
\end{align}
with the boundary condition $\nabla d \cdot \ve{n} = 0$. Here, in order to ensure $\dot{d}\geq0$, the history variable \cite{miehe_phase_2010} $\max_t\psie$ is applied. 

\subsubsection*{Fatigue variable}

The fatigue variable $D$ is a local variable computed at each material point with the \textit{Local Strain Approach} (LSA) \cite{seeger_grundlagen_1996}. Starting point is the von {Mises} equivalent stress $\sigma$ defined by the linear-elastic material model. It is then projected to the cyclic stress-strain curve (CSSC) assuming constant strain energy $\sigma \varepsilon = \sigma^*\varepsilon^*$, yielding a virtual stress-strain pair $(\sigma^*,\varepsilon^*)$ \cite{neuber_theory_1961}. In this way, a virtual stress-strain history considering cyclic plastic material behaviour can be determined for the complete loading sequence. This stress-strain path is segmented into hysteresis loops. The damaging effect of each loop $i$ is evaluated by a damage parameter \cite{smith_stressstrain_1970}
\begin{equation}
    P_{\mathrm{SWT},i} = \sqrt{(\sigma_{\mathrm{a},i}^*+\sigma_{\mathrm{m},i}^*)\,\varepsilon_{\mathrm{a},i}^*\,E}
\end{equation}
according to its stress amplitude $\sigma_\mathrm{a}$, mean stress $\sigma_\mathrm{m}$ and strain amplitude $\varepsilon_\mathrm{m}$. The damage parameter is then compared to a strain {Wöhler} curve, yielding a fatigue contribution $\Delta D_i$ for this single load cycle. Fatigue contributions from all passed load cycles can then be accumulated linearly to the fatigue variable $D =\sum_i \Delta D_i$. 

{The LSA represents a simplified way to include cyclic plasticity into the model, due to the revaluation of the stress-strain path to the CSSC with the Neuber rule. Please see \cite{seiler_efficient_2020} for a detailed discussion on this simplification.}
CSSC and strain {Wöhler} curve are material characteristics determined from standardised experiments, here described in Sec.~\ref{sec:cyclicpar}. 
Please see \cite{seiler_efficient_2020,seiler_phasefield_2021a} for a more detailed description of the integration of the Local Strain Approach in the phase-field fatigue model as well as its implementation in a material routine of a finite element code.
 
\subsection{Extension to transversal anisotropy}

\begin{figure} [t] 
	\def\svgwidth{\linewidth}\small{
		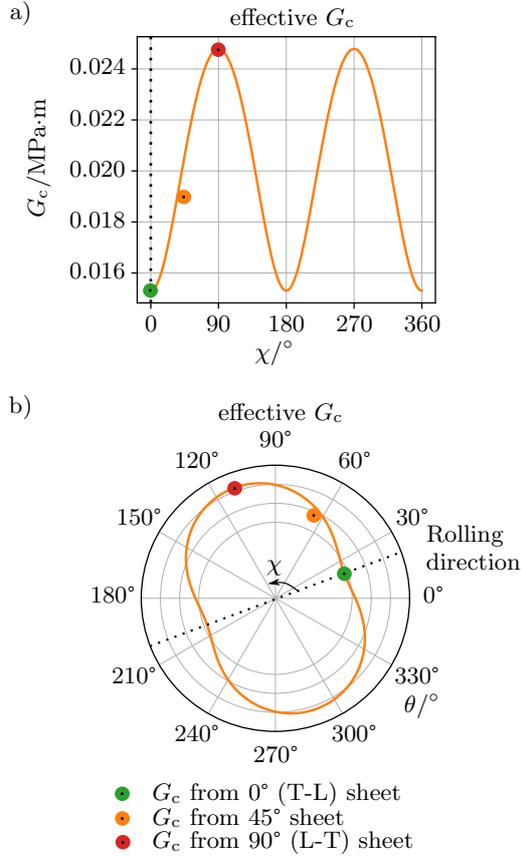}
	\caption{Effective fracture toughness $\Gc$ for anisotropic phase-field model according to  \cite{teichtmeister_phase_2017}, \textbf{(a)} displayed over angle from rolling direction $\chi$. \textbf{(b)} Polar plot includes arbitrary rolling direction of sheet material. Effective $\Gc$ was fitted to experimental results for angle to rolling direction $\theta=\{0^\circ,45^\circ,90^\circ\}$.}
	\label{fig:Gc}
\end{figure}

The anisotropy due to the process-induced elongation of the grains and oriented texture in the aluminium material could manifest in three ways: 
\begin{itemize}
    \item Anisotropic \textit{elastic} properties
    \item Ansiotropic static \textit{fracture resistance}, i.\,e. direction dependent fracture toughness $\Gc$
    \item Anisotropic \textit{fatigue resistance}, i.\,e. direction dependent CSSC and strain Wöhler curves
\end{itemize}
All three are studied in experiments described in Sec.~\ref{sec:exppar} for the given aluminium AA2024-T351. It turns out that for elastic and fatigue parameters, no significant direction dependence can be detected. Therefore, only the fracture toughness will be modelled as anisotropic. However, the extension to anisotropic elastic behaviour is straightforward. An approach to determine the fatigue variable given direction dependent fatigue parameters is described in \ref{app:anisofat}.

\subsubsection*{Anisotropic crack surface density}

The 2\,mm thin aluminium sheets are simulated in 2D with a plane stress state. Given the 2D problem and  only one existing preferred direction -- the rolling direction -- the material behaviour can be classified as \textit{transversely isotropic}. The approach adopted here to include transversely isotropic fracture resistance into the phase-field model was described first by {Teichtmeister} et al. \cite{teichtmeister_phase_2017} for static fracture. They introduce a structural tensor $\mathbf{A}$ into the gradient term of the crack surface density
\begin{equation} \label{eq:anisogam}
	\gamma_\mathrm{aniso}(d,\nabla d) = \frac{d^2}{2\ell} + \frac{\ell}{2}\nabla d \cdot \mathbf{A} \cdot \nabla d.
\end{equation}
The structural tensor is a tensor of second order and is defined by a direction vector $\mathbf{a}$, which is the rolling direction. The extent of anisotropy is characterised by the anisotropy parameter $\beta$, which has to be calibrated to represent the direction-dependent variation in crack resistance of the material. With both, the structural tensor is defined as  
\begin{equation} \label{eq:structtens}
	\mathbf{A} = \mathbf{I} + \beta(\mathbf{a}\otimes\mathbf{a}).
\end{equation}
Given the rolling direction is described by angle $\theta$, the direction vector is
\begin{equation}
	[a_l] = \left[ \begin{array}{c}
		\cos\theta \\
		\sin\theta \\
		0
	\end{array}
	 \right].
\end{equation}
Consequently, the structural tensor is
\begin{equation}
	[A_{lk}] = \left[ \begin{array}{ccc}
		1+\beta\cos^2\theta & \beta\sin\theta\cos\theta & 0 \\
		\beta\sin\theta\cos\theta & 1 + \beta\sin^2\theta & 0 \\
		0 &  0 & 1
	\end{array}
	\right].
\end{equation}
The resulting effective fracture toughness depending on the direction according to \cite{teichtmeister_phase_2017} is displayed in Fig.~\ref{fig:Gc}.

\subsubsection*{Governing equations}


Implementing the anisotropic crack surface density (\ref{eq:anisogam}) in the total energy functional (\ref{eq:Pi}) and repeating the variational principle, (\ref{eq:varprin}) yields the modified weak form of the phase-field problem
\begin{align} \label{eq:weakd}
	\delta_d \Pi= &\inte{\calB}{\left[g'(d)\psie_+(\epse)+{\alpha(D)}\frac{\Gc}{\ell}d  \right] \delta d}{v}  \nonumber \\
	& + \inte{\calB}{ {\alpha(D)}\Gc\ell \, \nabla d\cdot \mathbf{A} \cdot  \delta\nabla d }{v} = 0.
\end{align}


Subsequently, the evolution equation for the phase-field variable can be identified as
\begin{align} \label{eq:ev}
	2(1-d)\max_{\tau\in[0,t]}\psie(\tau) & = \Gc{\alpha(D)}\left(\ell\mathbf{A}\colon\Delta d -\frac{d}{\ell} \right) \nonumber\\ 
	&  + \, \Gc\ell\nabla d\cdot \mathbf{A}\cdot\nabla\alpha(D)
\end{align}
with the boundary condition
\begin{equation}
	\nabla d \cdot\mathbf{A}\cdot \ve{n} = 0.
\end{equation}

%

\section{Experiments for parameter identification} 
\label{sec:exppar}

Input parameters for the phase-field fatigue model are determined from various experiments.
The same material is used for the cyclic characterisation as for the fatigue crack growth experiments. Sheet material and properties are identical, see Sec.~\ref{sec:fatfrac}. Only CT specimens are cut from 25 mm thick plate material of AA2024-T351.

\subsection{Elastic parameters}

\begin{figure} [t] 
	\def\svgwidth{\linewidth}\small{
		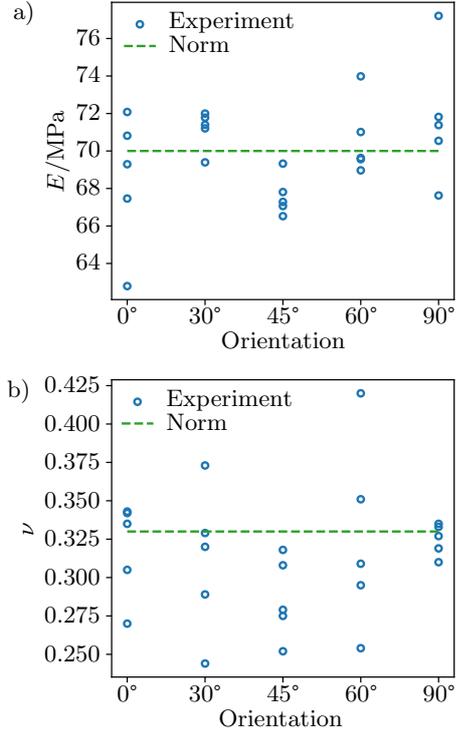}
	\caption{ \textbf{(a)} {Young}'s modulus and \textbf{(b)} {Poisson}'s ratio tested on tensile specimen cut from sheets with varying angle to rolling direction. Since no clear dependency on the angle is apparent, a constant value is set as recommended in norm SEP 1240 \cite{sep1240_testing_2006}. }
	\label{fig:elastic}
\end{figure}

Monotonic tensile tests with dog-bone shaped specimen cut from the aluminium sheet 0°, 30°, 45°, 60° and 90° angle w.\,r.\,t. the rolling direction were carried out. The determined {Young}'s modulus and {Poisson}'s ratio are displayed in Fig.~\ref{fig:elastic}. No clear dependency of both values on the orientation becomes apparent. Therefore, for the simulations, the standardised values of $E=70$~MPa and $\nu=0.33$ are used instead. This is also in accordance with norm SEP 1240 \cite{sep1240_testing_2006}, which recommends using standardised values for elastic parameters due to the scatter in most experiments. The norm is intended for steel material, though.

\subsection{Cyclic parameters}
\label{sec:cyclicpar}

In order to determine the fatigue variable $D$ with the Local Strain Approach, cyclic stress-strain curves (CSSC) and strain {Wöhler} curves are needed as an input for the model. Both are determined for dog-bone shaped specimens cut from the aluminium sheet in 0°, 45° and 90° angle to the rolling direction. Tests are performed according to SEP 1240 \cite{sep1240_testing_2006}. Even though intended for steel material, this norm has led to good results in the past also for aluminium {\cite{merklein_influence_2011}}. Since the load ratio is $R_\varepsilon=-1$, i\,e. symmetric loading in tensile and compressive direction, a Teflon-coated support system against buckling was installed.

\begin{figure} [t] 
	\def\svgwidth{\linewidth}\small{
		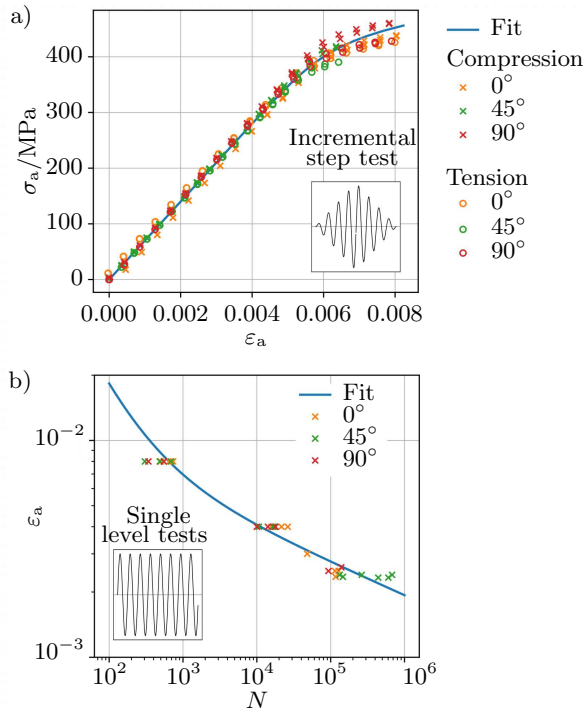}
	\caption{\textbf{(a)} cyclic stress-strain curve (CSSC) from incremental step test (IST) and \textbf{(b)} S-N  (Wöhler) curve recorded with single level tests. Tensile specimen cut from sheet material with varying angle to rolling direction were tested. The stress-strain tuples are evaluated seperately for tensile and compressive range. Since no dependency on neither angle nor loading direction is apparent, a fit (blue line) was done over all data sets for both curves.}
	\label{fig:WoehlerCSSC}
\end{figure}

\subsubsection*{Cyclic stress-strain curve}

The CSSC was determined with an Incremental Step Test (IST), i.\,e. with periodically rising and falling amplitudes \cite{landgraf_determination_1969}. The maximum strain amplitude was 0.8\,\% for 0° and 90° orientation and 0.65\,\% for 45° orientation. Tests for all orientations are evaluated separately for tensile and compressive range. As customary, the stress-strain characteristic after half the lifetime is selected for the fit. Results are displayed in Fig.~\ref{fig:WoehlerCSSC}b. Obviously, the stress-strain behaviour depends on neither the sheet orientation nor the loading direction. Therefore, all data is used for a single fit. The cyclic-stress strain curve according the the Ramberg-Osgood ansatz \cite{ramberg_description_1943}
\begin{equation} \label{eq:CSSC}
    \varepsilon_\mathrm{a} = \frac{\sigma_\mathrm{a}}{E} + \left( \frac{\sigma_\mathrm{a}}{K'} \right)^{1/n'}
\end{equation}
is parametrised with $K'=750$\,MPa and $n'=0.0728$, and is shown in blue in the graph. 

Here, we stick to the {usual} realisation of the Local Strain approach with only one CSSC. However, should the material show significant deviations in either tensile/compressive range or a drastic change of the behaviour over the specimens' liftime, a modification of the Local Strain approach by {Kühne} et al. \cite{kuhne_fatigue_2018} can be implemented. {It} allows for a tension-compression asymmetry and a transient evolution of the stress-strain behaviour. 

\subsubsection*{Strain Wöhler curve}

The strain {Wöhler} curve was recorded with single level tests with various strain amplitudes. Results are shown in Fig.~\ref{fig:WoehlerCSSC}a. Since no dependence on the rolling direction is apparent, only one single fit using all data was done. The fit yielded the parameters of the strain {Wöhler} curve
\begin{equation} \label{eq:SWC}
    \varepsilon_\mathrm{a} = \frac{\sigma'_\mathrm{f}}{E} \,(2N)^b + \varepsilon'_\mathrm{f}\,(2N)^c
\end{equation}
as \mbox{$\sigma'_\mathrm{f}=1206$\,MPa}, \mbox{$\varepsilon'_\mathrm{f}=0.9975$}, \mbox{$b=-0.152$} and \mbox{$c=-0.860$} and is marked in blue in Fig.~\ref{fig:WoehlerCSSC}a.

\subsection{Fracture toughness}

\begin{figure} [t] 
    \centering
	\def\svgwidth{0.8\linewidth}\small{
		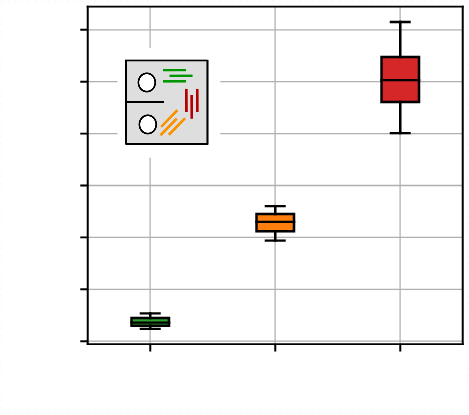}
	\begin{tabular}{l|ll}
	    Orientation & $K_{Ic}$ & $\Gc$ \\ \hline
	    0° (T-L)  & 34.7 MPa\,$\sqrt{\mathrm{m}}$ & 0.0153 MPa\,m \\ 
	    45°   & 38.6 MPa\,$\sqrt{\mathrm{m}}$ & 0.0189 MPa\,m \\
	    90° (L-T)  & 44.1 MPa\,$\sqrt{\mathrm{m}}$ & 0.0248 MPa\,m
	\end{tabular}
	\caption{Measured fracture toughness $K_{Ic}$ values for varying angle between rolling direction an crack $\chi$. Mean values displayed in table below, as well as corresponding $\Gc$. }
	\label{fig:KIc}
\end{figure}

The fracture toughness tests were carried out in accordance to ASTM E1820-01 \cite{astme1820-01_standard_2001}. In total twelve CT specimens, see Fig.~\ref{fig:specimen}b, were tested in three directions, with rolling directions as indicated in the top-left sketch in Fig.~\ref{fig:KIc}. The specimen geometry was $W= 50$\,mm, $a_0 =12.5$\,mm according to the norm,  with a thickness of $t=25$\,mm ensuring mostly plane strain conditions. {In the phase-field modelling community, $\Gc$ is treated as a material parameter, not a thickness-dependent structural parameter. Plane strain conditions are therefore chosen in order to obtain a parameter characterising the behaviour of material, ruling out influences of thickness and boundary effects as far as possible.
}

Tests were performed on a standard uniaxial servo hydraulic test rig. To determine the crack length during fatigue precracking, the direct current potential drop (DCPD) was used. Starting from an initial V-formed Chevron notch, a fatigue precrack was initialized with a constant $\Delta K =10\,\mathrm{MPa\sqrt{m}}$ and cyclic loading until an $a/W$ ratio of 0.45\dots0.5 was met. The test procedure for determing $K_{Ic}$ was displacement-controlled with a displacement rate of 1\,mm/minute. During this test sequence, the crack mouth opening displacement (CMOD) as well as the applied load $F$ were recorded. $K_{Ic}$ was determined according to the secant method based on the load vs. CMOD curve. 
Fig.~\ref{fig:KIc} shows the measured plane strain fracture toughness $K_{Ic}$ related to the specimen orientation, as well as the corresponding critical energy release rate commonly used in phase-field models 
\begin{equation}
    \Gc = \frac{1-\nu^2}{E}K_{Ic}.
\end{equation}
With a mean fracture toughness of $34.76\,\mathrm{MPa\sqrt{m}}$ the $\chi=0$°  orientation has the lowest fracture toughness and the highest value of $44.11\,\mathrm{MPa\sqrt{m}}$ can be measured for specimen in $\chi=90$°. The experimentally determined values for $\chi=90$° and $\chi=0$° are comparable to references given in literature \cite{_inc_,bucci_selecting_1979}, whereas no references exist for fracture toughness in 45° orientation.

The mean value of each orientation was used to fit the anisotropy parameter of the phase-field model according to eq.~(\ref{eq:structtens}), yielding $\beta=1.625$. For an illustration of the calibrated effective fracture toughness, see Fig.~\ref{fig:Gc}.

\section{Numerical results}

In the following, the developed model is used to simulate the FCG according to the experiments with MT specimen described in Sec.~\ref{sec:fatfrac}. For the elastic, cyclic and fracture parameters, the values determined experimentally in Sec.~\ref{sec:exppar} are used, if not stated otherwise. { All model parameters are summarised in Tab.~\ref{tab:paras}.} The MT geometry is meshed with quadratic elements. The mesh is refined in the relevant areas with a minimum mesh size of $h_{\min}=0.3$\,mm.  The coupled problem is solved with a staggered approach with convergence control.
Viscous regularisation with a viscous damping parameter $\eta$ is applied in order to compensate for convergence issues due to fast-developing cracks. See \cite{kalina_overview_2023} for a general phase-field fatigue model structure with viscous regularisation. {Here, the viscous damping parameter is chosen according to preliminary studies with static and cyclic loading, precluding its influence on the results.}

\begin{table}[]
    \centering
    \begin{tabular}{c|c} 
         Elastic &  $E=70$ MPa, $\mu=0.33$\\ \hline
         Fracture & $\Gc=0.0153$ MPa m, $\ell=0.5$ mm, \\
         & $\eta=1\cdot10^{-9}$, $\beta=1.625$ \\ \hline
         Cyclic stress- & $K'=750$\,MPa, $n'=0.0728$\\
         strain curve & \\ \hline
         Strain Wöhler & $\sigma'_\mathrm{f}=1206$\,MPa, $\varepsilon'_\mathrm{f}=0.9975$, \\
         curve & $b=-0.152$, \mbox{$c=-0.860$} \\ \hline
         Fatigue function & $\alpha_0=0.001$, $\xi=1450$ 
    \end{tabular}
    \caption{{Parameters of phase-field fatigue model.}}
    \label{tab:paras}
\end{table}

\subsection{Numerical studies of static loading}

Firstly, the MT specimen is loaded with a \textit{static} load in order to study the effect of the anisotropic crack surface density without the influence of fatigue damage accumulation. This is a purely numerical experiment, it is not compared to experimental data. The characteristic length is chosen to be $\ell=1$\,mm, yielding a ratio $h_{\min}/\ell \approx1/3$, surpassing the recommendation \cite{miehe_thermodynamically_2010} of $h_{\min}/\ell <1/2$. A total displacement of 2\,mm is applied. The viscosity parameter is set to $\eta=1\cdot10^{-6}$. Apart from the experimentally determined anisotropy parameter $\beta=1.625$, $\beta=0.4$ is tested for comparison.

\begin{figure*} [t] 
	\def\svgwidth{\textwidth}\small{
		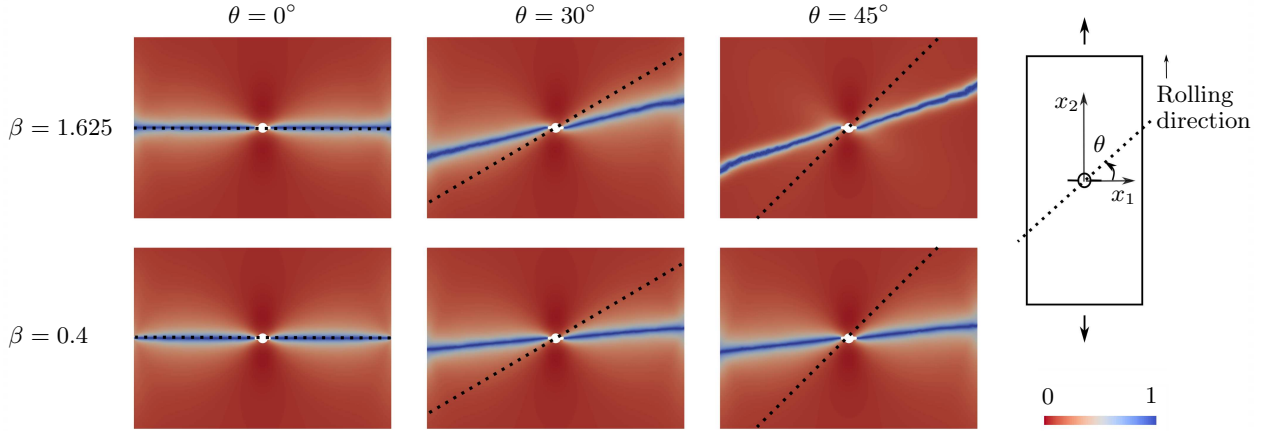}
	\caption{Simulation of \textit{static} fracture with anisotropic phase-field model. Shown is a contour plot of the phase-field variable $d$. $\beta$ is the anisotropy parameter. The bigger $\beta$, the stronger the crack inclines towards the rolling direction, as anisotropy in the fracture toughness is more pronounced.
	}
	\label{fig:staticMT}
\end{figure*}

The phase-field plots are shown in Fig.~\ref{fig:staticMT}. Only the center part of the MT specimen is displayed. The rolling direction, which was varied between $\theta=0$° to 45° from the horizontal direction, is marked with a dashed black line. As expected, for horizontal rolling direction ($\theta=0$°), the crack path is horizontal as well, independent of the anisotropy parameter $\beta$. Crack inclination increases with steeper rolling direction angles. The effect is stronger for a higher anisotropy parameters $\beta$. Still, the original difference in effective $\Gc$ between 0° and 90° direction of about 30\,\% is still quite low compared to other typical anisotropic materials, like e.\,g. fibre reinforced composites. This explains that the crack does not follow the rolling direction (which is the direction of lowest effective fracture resistance) strictly. See \cite{teichtmeister_phase_2017} for a study on the degree of anisotropy in anisotropic static phase-field models.

\subsection{Parametrisation of cyclic model}
\label{sec:param}

The cyclic MT experiment used for fitting the model is the $\theta=0^\circ$ orientation with $R=0.1$ and a constant cyclic load amplitude of $F_{\max}=15$\,kN. The damping parameter is set to $\eta=1\cdot10^{-9}$ and the characteristic length to $\ell=0.5$\,mm. With all other parameters coming from the experiments in Sec.~\ref{sec:exppar}, only the parameters of the fatigue degradation $\alpha(D)$ function remain to be fitted.

\begin{figure} [p] 
	\def\svgwidth{\linewidth}\small{
		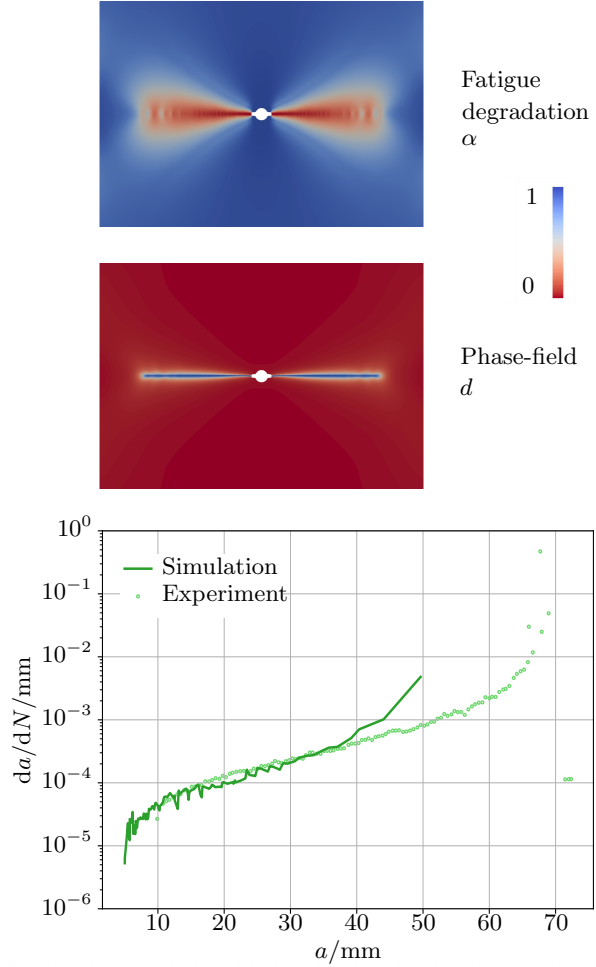}
	\caption{Fit of phase-field fatigue model to MT experiment with $\theta=0^\circ,R=0.1$. Fit yielded parameters of fatigue degradation function $\alpha_0=0.001, \xi=1450$. Displayed are the fatigue degradation $\alpha$ and the phase-field at the end of the simulation, as well as the fatigue crack growth rate plotted over crack length $a$.
	}
	\label{fig:fit}
\end{figure}

The results are summarised in Fig.~\ref{fig:fit}. {Please note that the crack propagation rate is now plotted over the crack length $a$ instead of $\Delta K$ as in Fig.~\ref{fig:exp_results} in order to allow for a better comparison with phase-field contour plots of the crack.} The figure shows the distribution of the fatigue degradation function $\alpha$ and the phase-field $d$ at the end of the simulation. They underline how the degradation of the fracture resistance $\Gc$ has allowed the phase-field crack grow at a sub-critical, but repetitive load. The calibration to the experimental FCG rate yielded the parameters $\alpha_0=0.001$ and $\xi=1450$, as displayed in the lower graph. The calibrated model reproduces the experimental data fairly well in the first two thirds of crack length $a$. At a later stage, the simulation is not able to capture the transition to unstable, static residual fracture at about $a=50$\,mm almost at the outer boundary of the specimen.

\subsection{Validation}

\begin{figure} [p] 
	\def\svgwidth{\linewidth}\small{
		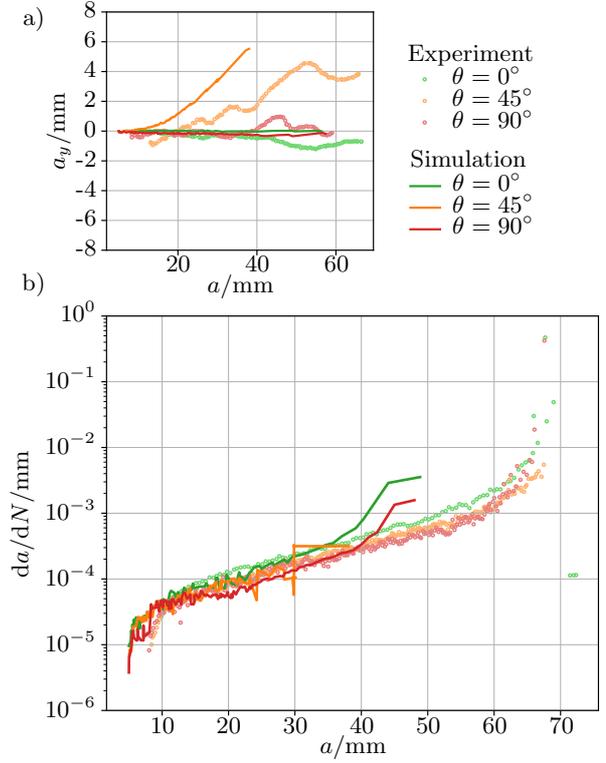}
	\caption{Simulation of fatigue crack growth in MT specimen with different angles to rolling direction $\theta$ and $R=0.1$, see Fig.~\ref{fig:staticMT} for geometry. Fit of $\theta=0^\circ$ was used for all simulations. \textbf{(a)} Simulation reproduces horizontal crack path for $\theta=0^\circ,90^\circ$ and inclination for $\theta=45^\circ$. \textbf{(b)} In experiments and simulations, crack propagation rate is the highest for $\theta=0^\circ$ and the lowest for $\theta=90^\circ$.
	}
	\label{fig:rollingdir}
\end{figure}

For validation, the fit is used to simulate the FCG experiments that were not used for parameter identification. All parameters remain identical to those stated in Sec.~\ref{sec:param}. First, the orientation of the sheet material is varied. Fig.~\ref{fig:rollingdir} shows a) the resulting crack path and b) the fatigue crack propagation rate. Consistent with the experimental results, both the 0° and 90° orientation lead to a straight crack. The crack in the sheet with 45° orientation, on the other hand, inclines to about 8.5°. Apart from the jagged experimental crack geometry, the simulation yields a similar crack path compared to the experiment. The crack propagation rates differ -- as in the experiments -- only slightly. The correct order of crack propagation rates, with the 0° orientation being the fastest and 90° the slowest -- is reproduced correctly.

\begin{figure} [p] 
	\def\svgwidth{\linewidth}\small{
		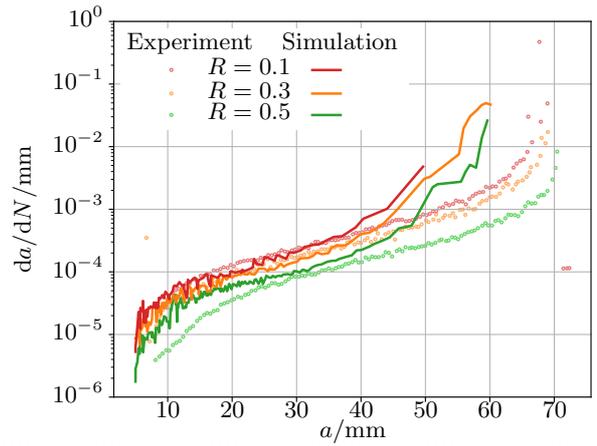}
	\caption{Fatigue crack growth for different load ratios $R$ and $\theta=0$°. Fit of $R=0.1$ was used in all simulations. Experiments' tendency of lower $R$ to have the highest crack propagation rate is reproduced.
	}
	\label{fig:r-ratio}
\end{figure}

The final study addresses the $R$-ratio dependence. As shown in Fig.~\ref{fig:r-ratio}, additionally, $R=0.3$ and $R=0.5$ are tested with the model fitted to the $R=0.1$ experiments. The maximum load $F_{\max}=15$\,kN is kept constant, while $F_{\min}$ is increased to 4.5\,kN and 7.5\,kN, thereby reducing the force amplitude. The simulation correctly reproduces the tendency of higher $R$-ratios to lead to lower fatigue crack propagation rates.

{ The results do not match the experiments exactly, though. For one thing, this is owed to doing the simulation in 2D. Although the sheet is quite thin with a thickness of only 2~mm, the plane stress state is only an approximation of the present stress state. Variation of the grain structure over the thickness, boundary effects, curved crack fronts and tilting of the crack front angle can be observed for fatigue fracture in this material. Moreover, as a ductile material, aluminium can form significant plastic zones, influencing the stress state at the crack tip.  The LSA only uses a simplified plasticity model based on the cyclic stress-strain curve, see \cite{seiler_efficient_2020} for details.
Both the extension to 3D and to elastic-plastic material modelling are out of this paper, though, and subject of future work.}

\section{Conclusions}

FCG behaviour in rolled aluminium sheets shows a dependency of the crack path w.\,r.\,t. the sheet rolling direction, as well as a growth rate that decreases with increasing load ratio. The cyclic crack growth is modelled with the phase-field method due to its superior ability to model arbitrary crack paths in a straightforward way. The fatigue variable of the model is based on the Local Strain Approach. An extension to anisotropic material behaviour is proposed by introducing a structural tensor in the regularised crack surface density. Thereby, the effective fracture toughness is interpolated. 

The model is parameterised for AA2024-T351 sheet material using experimentally determined fracture toughness values in three different orientations. Elastic and cyclic properties do not exhibit a direction dependency. The parameterised model reproduces the influence of the load ratio on the fatigue crack propagation rate as well as the crack angle depending on the sheet orientation.

So far, the simulation of the 2\,mm thick sheets are performed with a 2D model assuming a plane stress state. This raises the question of the correct choice of the fracture toughness $\Gc$. In the phase-field community, $\Gc$ is mostly seen as a material parameter. In accordance with that, this paper determines $\Gc$ with thick specimen, approximating a plane strain state, as this yields a universal material value. If thin sheets are used for the determination, the parameter turns out to be thickness-dependent. Therefore, it could be appropriate to view $\Gc$ as a structural parameter instead, which has to be calibrated to the thickness of the specimen -- if 2D simulations are used.

Although the model is able to reproduce the main characteristics of crack propagation behaviour, certain phenomena cannot be covered with a 2D model. This includes three dimensional effects as variation of the crack front, for example crack tunnelling or slant crack growth, that affect crack closure behaviour and therefore lead to variation in crack propagation rates. Hence, the development of 3D models is inevitable in the future. The model presented here uses a simplified cyclic plasticity model for damage calculation. However, the effect of the plastic zone should be studied in more detail, using a proper elastic-plastic material model, especially if higher loadings $\Delta K$ are investigated.

\section*{Acknowledgements}

This work was supported by the Deutsche Forschungsgemeinschaft (DFG) via the project \textit{Experimental analysis and phase-field modelling of the interaction between plastic zone and fatigue crack growth in ductile materials under complex loading} (grant numbers KA 3309/12-1, {BR 6259/2-1}). Furthermore, funding came from the Federal Ministry for Economic Affairs and Climate Action on the basis of a decision by the German Bundestag, within the aerospace program LuFo-VI of the project \textit{Untersuchung der Prozess-Struktur-Eigenschaftsbeziehungen des Rollprofilierens auf die Performanceverbesserung für Blechbauteile aus Aluminium} (Funding ID 20W2102D). The authors are grateful to the Centre for Information Services and High Performance Computing (ZIH) TU Dresden for providing its facilities for high throughput calculations.

\section*{Highlights} 

\begin{enumerate}
    \item Development of the first phase-field model for fatigue fracture in anisotropic media
	\item {Acquisition of set of direction-dependent material parameters for a AA2024-T351 aluminium material, including fracture toughness and cyclic parameters}
	\item {Parametrization of the model for anisotropic aluminum sheet material}
	\item Reproduction of experimentally observed effects in the crack path and crack growth rate
\end{enumerate}

\appendix

\section{Anisotropic fatigue damage calculation}
\label{app:anisofat}

The cyclic material behaviour of AA2024-T351 is not direction dependent, as shown in Sec.~\ref{sec:cyclicpar}. However, should the desired material show a cyclic anisotropy, the following anisotropic fatigue concept can be implemented. 

\begin{figure*} [t] 
	\def\svgwidth{\textwidth}\small{
		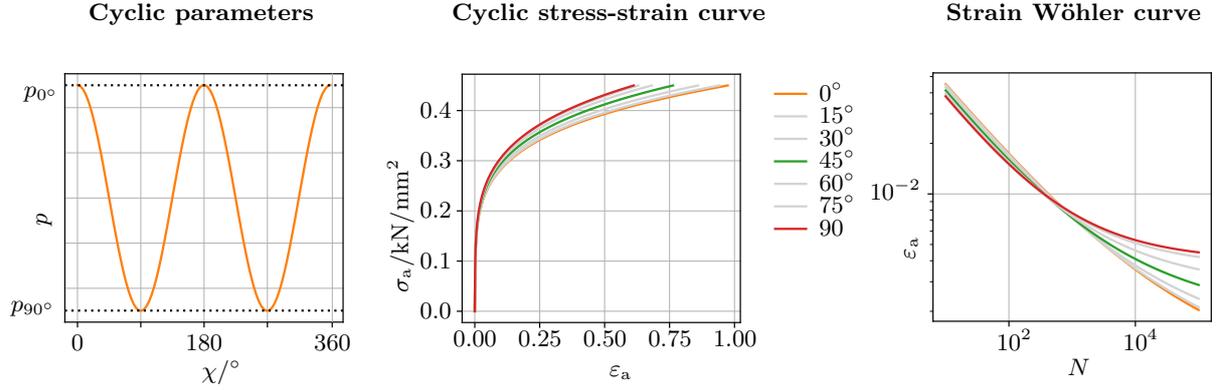}
	\caption{ Extension of model to anisotropic cyclic parameters: All parameters of cyclic stress-strain curve and strain Wöhler curve, represented by $p$, are interpolated with sine function between their value in and perpendicular to rolling direction. This leads to an array of curves.}
	\label{fig:directiondep}
\end{figure*}

The parameters of the CSSC (\ref{eq:CSSC}) $K'$ and $n'$ and the strain {Wöhler} curve (\ref{eq:SWC}) $\sigma'_\mathrm{f},\varepsilon'_\mathrm{f},b$ and $c$ are determined for 0° and 90° orientation individually. All parameters -- here represented by $p$ -- are then interpreted as direction dependent parameters $K'(\chi),n'(\chi),\dots$ and interpolated over $\chi$ with the cosine ansatz
\begin{equation}
	p(\chi) = \frac{p_{0^\circ}-p_{90^\circ}}{2}\cos(2\chi) + p_{0^\circ} + \frac{p_{90^\circ}-p_{0^\circ}}{2},
\end{equation}
as shown in Fig.~\ref{fig:directiondep} on the left. $p_{0^\circ}$  and $p_{90^\circ}$ are the experimental values of the parameter in 0° and 90° orientation, respectively. The angle $\chi$ is measured from the rolling direction. This yields a family of curves for each the CSSC and the strain {Wöhler curve}, see Fig.~\ref{fig:directiondep}. The interpolation should be checked with experimentally determined curves in between 0° and 90°. In this way, the material resistance against fatigue can be described for arbitrary directions in the 2D plane.

\begin{figure*} [t] 
	\def\svgwidth{\linewidth}\small{
		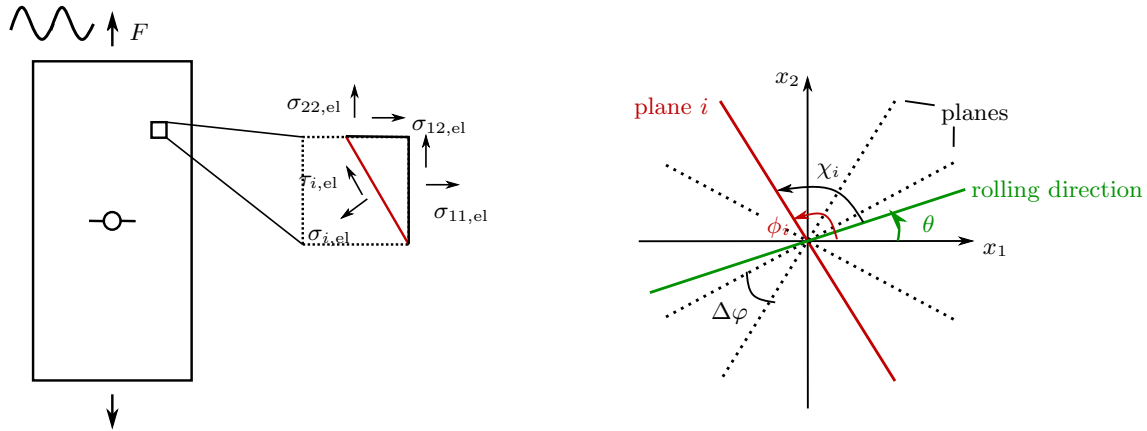}
	\caption{ Extension of model to anisotropic cyclic parameters: Damage concept at each material point involves critical plane approach. 2D space is sampled by finite number of planes $i$, for which damage is calculated individually, considering stress state and cyclic parameters (see Fig.~\ref{fig:directiondep}) for that direction. Maximum of damages of all directions finally enters fatigue degradation function.}
	\label{fig:critplane}
\end{figure*}

The fatigue variable at each material point is then determined with the critical plane approach \cite{radaj_ermudungsfestigkeit_2007}. Thereby, the 2D spaces is sampled by a discrete number of planes at each point, see Fig.~\ref{fig:critplane}. For each plane, the stress state and the materials' resistance is determined in the particular direction. They are the input to the Local Strain Approach, instead of e.\,g. the equivalent stress in the standard procedure. Finally, the fatigue variable is then determined as the maximum of all fatigue values in the set of directions at this particular point.
The quality of the result depends, among others, on the amount of planes used in the sampling.

\bibliography{mybibfile}	

\begin{thebibliography}{10}
\expandafter\ifx\csname url\endcsname\relax
  \def\url#1{\texttt{#1}}\fi
\expandafter\ifx\csname urlprefix\endcsname\relax\def\urlprefix{URL }\fi
\expandafter\ifx\csname href\endcsname\relax
  \def\href#1#2{#2} \def\path#1{#1}\fi

\bibitem{tavares_overview_2017}
S.~M.~O. Tavares, P.~M. S.~T. de~Castro, An overview of fatigue in aircraft
  structures: {Overview} of {Fatigue} in {Aircraft} {Structures}, Fatigue \&
  Fracture of Engineering Materials \& Structures 40~(10) (2017) 1510--1529.
\newblock \href {http://dx.doi.org/10.1111/ffe.12631}
  {\path{doi:10.1111/ffe.12631}}.

\bibitem{schijve_fatigue_2003}
J.~Schijve, Fatigue of structures and materials in the 20th century and the
  state of the art, International Journal of Fatigue 25~(8) (2003) 679--702.
\newblock \href {http://dx.doi.org/10.1016/S0142-1123(03)00051-3}
  {\path{doi:10.1016/S0142-1123(03)00051-3}}.

\bibitem{astme647-15_standard_2015}
A.~E647-15, Standard {Test} {Method} for {Measurement} of {Fatigue} {Crack}
  {Growth} {Rates}, Tech. rep., ASTM International, West Conshohocken (2015).
\newblock \href {http://dx.doi.org/10.1520/E0647-05}
  {\path{doi:10.1520/E0647-05}}.

\bibitem{newman_fastran2_1992}
J.~C. Newman, Jr., {FASTRAN}-2: {A} fatigue crack growth structural analysis
  program, NASA STI/Recon Technical Report N 92 (1992) 30964, aDS Bibcode:
  1992STIN...9230964N.

\bibitem{_nasgro_2010}
{NASGRO}: {Fracture} mechanics and fatigue crack growth analysis software
  (2010).

\bibitem{sagrillo_elastic_2022}
C.~N. Sagrillo, L.~Shimizu, V.~K. Goyal, Elastic–{Plastic} {Fracture}
  {Mechanics} {Guidance} and {Analysis} {Validation}, Journal of Spacecraft and
  Rockets 59~(6) (2022) 1869--1884.
\newblock \href {http://dx.doi.org/10.2514/1.A35387}
  {\path{doi:10.2514/1.A35387}}.

\bibitem{Noroozi2005}
A.~H. Noroozi, G.~Glinka, S.~Lambert, A two parameter driving force for fatigue
  crack growth analysis, Vol.~27, 2005, pp. 1277--1296.
\newblock \href {http://dx.doi.org/10.1016/j.ijfatigue.2005.07.002}
  {\path{doi:10.1016/j.ijfatigue.2005.07.002}}.

\bibitem{Mikheevskiy2015}
S.~Mikheevskiy, S.~Bogdanov, G.~Glinka, Analysis of fatigue crack growth under
  spectrum loading - the unigrow fatigue crack growth model, Theoretical and
  Applied Fracture Mechanics 79 (2015) 25--33.
\newblock \href {http://dx.doi.org/10.1016/j.tafmec.2015.06.010}
  {\path{doi:10.1016/j.tafmec.2015.06.010}}.

\bibitem{Bang2019}
D.~J. Bang, A.~Ince, L.~Q. Tang, A modification of unigrow 2-parameter driving
  force model for short fatigue crack growth, Fatigue and Fracture of
  Engineering Materials and Structures 42 (2019) 45--60.
\newblock \href {http://dx.doi.org/10.1111/ffe.12865}
  {\path{doi:10.1111/ffe.12865}}.

\bibitem{american1971damage}
W.~Elber, The significance of fatigue crack closure, in: Damage tolereance in
  aircraft structures, {ASTM} special technical publication, American Soc. for
  Testing and Materials, 1971, pp. 230--242.

\bibitem{needleman_analysis_1987}
A.~Needleman, V.~Tvergaard, An analysis of ductile rupture modes at a crack
  tip, Journal of the Mechanics and Physics of Solids 35~(2) (1987) 151--183.
\newblock \href {http://dx.doi.org/10.1016/0022-5096(87)90034-2}
  {\path{doi:10.1016/0022-5096(87)90034-2}}.

\bibitem{ritchie_no_1999}
R.~Ritchie, [{No} title found], International Journal of Fracture 100~(1)
  (1999) 55--83.
\newblock \href {http://dx.doi.org/10.1023/A:1018655917051}
  {\path{doi:10.1023/A:1018655917051}}.

\bibitem{dursun_recent_2014}
T.~Dursun, C.~Soutis, Recent developments in advanced aircraft aluminium
  alloys, Materials \& Design (1980-2015) 56 (2014) 862--871.
\newblock \href {http://dx.doi.org/10.1016/j.matdes.2013.12.002}
  {\path{doi:10.1016/j.matdes.2013.12.002}}.

\bibitem{wei_influence_2014}
L.~Wei, Q.~Pan, H.~Huang, L.~Feng, Y.~Wang, Influence of grain structure and
  crystallographic orientation on fatigue crack propagation behavior of 7050
  alloy thick plate, International Journal of Fatigue 66 (2014) 55--64.
\newblock \href {http://dx.doi.org/10.1016/j.ijfatigue.2014.03.009}
  {\path{doi:10.1016/j.ijfatigue.2014.03.009}}.

\bibitem{xia_texture_2018}
P.~Xia, Z.~Liu, W.~Wu, Q.~Zhao, P.~Ying, S.~Bai, Texture {Effect} on {Fatigue}
  {Crack} {Propagation} {Behavior} in {Annealed} {Sheets} of an {Al}-{Cu}-{Mg}
  {Alloy}, Journal of Materials Engineering and Performance 27~(9) (2018)
  4693--4702.
\newblock \href {http://dx.doi.org/10.1007/s11665-018-3582-5}
  {\path{doi:10.1007/s11665-018-3582-5}}.

\bibitem{gilmour_influence_2004}
K.~Gilmour, A.~Leacock, M.~Ashbridge, The influence of plastic strain ratios on
  the numerical modelling of stretch forming, Journal of Materials Processing
  Technology 152~(1) (2004) 116--125.
\newblock \href {http://dx.doi.org/10.1016/j.jmatprotec.2004.03.013}
  {\path{doi:10.1016/j.jmatprotec.2004.03.013}}.

\bibitem{barlat_sixcomponent_1991}
F.~Barlat, D.~J. Lege, J.~C. Brem, A six-component yield function for
  anisotropic materials, International Journal of Plasticity 7~(7) (1991)
  693--712.
\newblock \href {http://dx.doi.org/10.1016/0749-6419(91)90052-Z}
  {\path{doi:10.1016/0749-6419(91)90052-Z}}.

\bibitem{seidt_plastic_2013}
J.~Seidt, A.~Gilat, Plastic deformation of 2024-{T351} aluminum plate over a
  wide range of loading conditions, International Journal of Solids and
  Structures 50~(10) (2013) 1781--1790.
\newblock \href {http://dx.doi.org/10.1016/j.ijsolstr.2013.02.006}
  {\path{doi:10.1016/j.ijsolstr.2013.02.006}}.

\bibitem{amstutz_effects_1997}
B.~E. Amstutz, M.~A. Sutton, D.~S. Dawicke, M.~L. Boone, Effects of mixed mode
  {I}/{II} loading and grain orientation on crack initiation and stable tearing
  in 2024-{T3} aluminum, ASTM STP 1296 (1997) 105--125.

\bibitem{johnston_fracture_2001}
W.~M. Johnston, J.~C. Newman, Fracture tests on thin sheet 2024-{T3} aluminum
  alloy for specimens with and without anti-buckling guides, Tech. rep. (2001).

\bibitem{bergner_new_2000}
F.~Bergner, A new approach to the correlation between the coefficient and the
  exponent in the power law equation of fatigue crack growth, International
  Journal of Fatigue 22~(3) (2000) 229--239.
\newblock \href {http://dx.doi.org/10.1016/S0142-1123(99)00123-1}
  {\path{doi:10.1016/S0142-1123(99)00123-1}}.

\bibitem{carvalhodacunha_influence_2017}
M.~Carvalho~da Cunha, M.~S. Fernandes~de Lima, The influence of laser surface
  treatment on the fatigue crack growth of {AA} 2024-{T3} aluminum alloy alclad
  sheet, Surface and Coatings Technology 329 (2017) 244--249.
\newblock \href {http://dx.doi.org/10.1016/j.surfcoat.2017.08.052}
  {\path{doi:10.1016/j.surfcoat.2017.08.052}}.

\bibitem{kocanda_probabilistic_2012}
D.~Kocańda, M.~Jasztal, Probabilistic predicting the fatigue crack growth
  under variable amplitude loading, International Journal of Fatigue 39 (2012)
  68--74.
\newblock \href {http://dx.doi.org/10.1016/j.ijfatigue.2011.03.011}
  {\path{doi:10.1016/j.ijfatigue.2011.03.011}}.

\bibitem{kuna_numerische_2010}
M.~Kuna, Numerische {Beanspruchungsanalyse} von {Rissen}: finite {Elemente} in
  der {Bruchmechanik} ; mit zahlreichen {Beispielen}, 2nd Edition, Aus dem
  {Programm} {Mechanik}, Vieweg + Teubner, Wiesbaden, 2010, oCLC: 845668915.

\bibitem{moes_finite_1999}
N.~Moës, J.~Dolbow, T.~Belytschko, A finite element method for crack growth
  without remeshing, International Journal for Numerical Methods in Engineering
  46~(1) (1999) 131--150.
\newblock \href
  {http://dx.doi.org/10.1002/(SICI)1097-0207(19990910)46:1<131::AID-NME726>3.0.CO;2-J}
  {\path{doi:10.1002/(SICI)1097-0207(19990910)46:1<131::AID-NME726>3.0.CO;2-J}}.

\bibitem{ambati_phasefield_2015}
M.~Ambati, T.~Gerasimov, L.~De~Lorenzis, Phase-field modeling of ductile
  fracture, Computational Mechanics 55~(5) (2015) 1017--1040.
\newblock \href {http://dx.doi.org/10.1007/s00466-015-1151-4}
  {\path{doi:10.1007/s00466-015-1151-4}}.

\bibitem{alessi_comparison_2018}
R.~Alessi, M.~Ambati, T.~Gerasimov, S.~Vidoli, L.~De~Lorenzis, Comparison of
  {Phase}-{Field} {Models} of {Fracture} {Coupled} with {Plasticity}, in:
  E.~Oñate, D.~Peric, E.~de~Souza~Neto, M.~Chiumenti (Eds.), Advances in
  {Computational} {Plasticity}, Vol.~46, Springer International Publishing,
  Cham, 2018, pp. 1--21.
\newblock \href {http://dx.doi.org/10.1007/978-3-319-60885-3_1}
  {\path{doi:10.1007/978-3-319-60885-3_1}}.

\bibitem{dammass_unified_2021}
F.~Dammaß, M.~Ambati, M.~Kästner, A unified phase-field model of fracture in
  viscoelastic materials, Continuum Mechanics and Thermodynamics 33~(4) (2021)
  1907--1929.
\newblock \href {http://dx.doi.org/10.1007/s00161-021-01013-3}
  {\path{doi:10.1007/s00161-021-01013-3}}.

\bibitem{scherer_assessment_2022}
J.-M. Scherer, S.~Brach, J.~Bleyer, An assessment of anisotropic phase-field
  models of brittle fracture, Computer Methods in Applied Mechanics and
  Engineering 395 (2022) 115036.
\newblock \href {http://dx.doi.org/10.1016/j.cma.2022.115036}
  {\path{doi:10.1016/j.cma.2022.115036}}.

\bibitem{kalina_overview_2023}
M.~Kalina, T.~Schneider, J.~Brummund, M.~Kästner, Overview of phase-field
  models for fatigue fracture in a unified framework, arXiv:2302.01396
  [cond-mat] (Feb. 2023).

\bibitem{seiler_efficient_2020}
M.~Seiler, T.~Linse, P.~Hantschke, M.~Kästner, An efficient phase-field model
  for fatigue fracture in ductile materials, Engineering Fracture Mechanics 224
  (2020) 106807.
\newblock \href {http://dx.doi.org/10.1016/j.engfracmech.2019.106807}
  {\path{doi:10.1016/j.engfracmech.2019.106807}}.

\bibitem{teichtmeister_phase_2017}
S.~Teichtmeister, D.~Kienle, F.~Aldakheel, M.-A. Keip, Phase field modeling of
  fracture in anisotropic brittle solids, International Journal of Non-Linear
  Mechanics 97 (2017) 1--21.
\newblock \href {http://dx.doi.org/10.1016/j.ijnonlinmec.2017.06.018}
  {\path{doi:10.1016/j.ijnonlinmec.2017.06.018}}.

\bibitem{johnson_calibrating_1965}
H.~H. Johnson, Calibrating the electric potential method for studying slow
  crack growth. 5.1 (1965) 442--445.

\bibitem{feddersen_discussion_1966}
C.~E. Feddersen, Discussion to "{Plane} strain crack toughness testing of high
  strength metallic materials", Tech. Rep. 410, ASTM International (1966).

\bibitem{melching_explainable_2022}
D.~Melching, T.~Strohmann, G.~Requena, E.~Breitbarth, Explainable machine
  learning for precise fatigue crack tip detection, Scientific Reports 12~(1)
  (2022) 9513.
\newblock \href {http://dx.doi.org/10.1038/s41598-022-13275-1}
  {\path{doi:10.1038/s41598-022-13275-1}}.

\bibitem{strohmann_automatic_2021}
T.~Strohmann, D.~Starostin‐Penner, E.~Breitbarth, G.~Requena, Automatic
  detection of fatigue crack paths using digital image correlation and
  convolutional neural networks, Fatigue \& Fracture of Engineering Materials
  \& Structures 44~(5) (2021) 1336--1348.
\newblock \href {http://dx.doi.org/10.1111/ffe.13433}
  {\path{doi:10.1111/ffe.13433}}.

\bibitem{jesus_influence_2022}
J.~Jesus, F.~Antunes, P.~Prates, R.~Branco, P.~Antunes, L.~Borrego, D.~Neto,
  Influence of specimen orientation on fatigue crack growth in 7050-{T7451} and
  2050-{T8} aluminium alloys, International Journal of Fatigue 164 (2022)
  107136.
\newblock \href {http://dx.doi.org/10.1016/j.ijfatigue.2022.107136}
  {\path{doi:10.1016/j.ijfatigue.2022.107136}}.

\bibitem{miehe_phase_2010}
C.~Miehe, M.~Hofacker, F.~Welschinger, A phase field model for rate-independent
  crack propagation: {Robust} algorithmic implementation based on operator
  splits, Computer Methods in Applied Mechanics and Engineering 199~(45) (2010)
  2765 -- 2778.
\newblock \href {http://dx.doi.org/https://doi.org/10.1016/j.cma.2010.04.011}
  {\path{doi:https://doi.org/10.1016/j.cma.2010.04.011}}.

\bibitem{seeger_grundlagen_1996}
T.~Seeger, Grundlagen für {Betriebsfestigkeitsnachweise}, Fundamentals for
  Service Fatigue-Strength Assessments),” Stahlbau Handbuch (Handbook of
  Structural Engineering), Stahlbau-Verlags-gesellschaft, Cologne 1 (1996)
  5--123.

\bibitem{neuber_theory_1961}
H.~Neuber, Theory of {Stress} {Concentration} for {Shear}-{Strained}
  {Prismatical} {Bodies} {With} {Arbitrary} {Nonlinear} {Stress}-{Strain}
  {Law}, Journal of Applied Mechanics 28~(4) (1961) 544 -- 550.
\newblock \href {http://dx.doi.org/http://dx.doi.org/10.1115/1.3641780}
  {\path{doi:http://dx.doi.org/10.1115/1.3641780}}.

\bibitem{smith_stressstrain_1970}
K.~N. Smith, T.~Topper, P.~Watson, A stress-strain function for the fatigue of
  metals, Journal of Materials 5 (1970) 767--778.

\bibitem{seiler_phasefield_2021a}
M.~Seiler, S.~Keller, N.~Kashaev, B.~Klusemann, M.~Kästner, Phase-field
  modelling for fatigue crack growth under laser shock peening-induced residual
  stresses, Archive of Applied Mechanics 91~(8) (2021) 3709--3723.
\newblock \href {http://dx.doi.org/10.1007/s00419-021-01897-2}
  {\path{doi:10.1007/s00419-021-01897-2}}.

\bibitem{sep1240_testing_2006}
S.~1240, Testing and {Documentation} {Guideline} for the {Experimental}
  {Determination} of {Mechanical} {Properties} of {Steel} {Sheets} for
  {CAE}-{Calculations}, {STAHL}-{EISEN}-{Prüfblätter} ({SEP}) des
  {Stahlinstitut} {VDEh} (2006).

\bibitem{merklein_influence_2011}
M.~Merklein, K.~Andreas, U.~Engel, Influence of machining process on residual
  stresses in the surface of cemented carbides, Procedia Engineering 19 (2011)
  252 -- 257.
\newblock \href
  {http://dx.doi.org/https://doi.org/10.1016/j.proeng.2011.11.108}
  {\path{doi:https://doi.org/10.1016/j.proeng.2011.11.108}}.

\bibitem{landgraf_determination_1969}
R.~W. Landgraf, J.~D. Morrow, T.~Endo, Determination of the cyclic
  stress-strain curve, Journal of materials~(1) (1969) 176--188.

\bibitem{ramberg_description_1943}
W.~Ramberg, W.~Osgood, Description of stress-strain curves by three parameters,
  NACA Technical Note 902.

\bibitem{kuhne_fatigue_2018}
D.~Kühne, C.~Guilleaume, M.~Seiler, P.~Hantschke, F.~Ellmer, T.~Linse,
  A.~Brosius, M.~Kästner, Fatigue analysis of rolled components considering
  transient cyclic material behaviour and residual stresses, Production
  Engineering\href {http://dx.doi.org/10.1007/s11740-018-0861-9}
  {\path{doi:10.1007/s11740-018-0861-9}}.

\bibitem{astme1820-01_standard_2001}
A.~E1820-01, Standard {Test} {Method} for {Measurement} of {Fracture}
  {Toughness}, Tech. rep., ASTM International, West Conshohocken (2001).
\newblock \href {http://dx.doi.org/10.1520/E1820-01}
  {\path{doi:10.1520/E1820-01}}.

\bibitem{_inc_}
Inc. {Material} {Data} {Sheet}: {Aluminum} 2024-{T4}; 2024-{T351}, Tech. rep.

\bibitem{bucci_selecting_1979}
R.~J. Bucci, G.~Nordmark, E.~A. Starke, Selecting {Aluminum} {Alloys} to
  {Resist} {Failure} by {Fracture} {Mechanisms}, Engineering Fracture
  Mechanics~(12.3) (1979) 407--441.

\bibitem{miehe_thermodynamically_2010}
C.~Miehe, F.~Welschinger, M.~Hofacker, Thermodynamically consistent phase-field
  models of fracture: {Variational} principles and multi-field {FE}
  implementations, International Journal for Numerical Methods in Engineering
  83~(10) (2010) 1273--1311.
\newblock \href {http://dx.doi.org/10.1002/nme.2861}
  {\path{doi:10.1002/nme.2861}}.

\bibitem{radaj_ermudungsfestigkeit_2007}
D.~Radaj, M.~Vormwald, Ermüdungsfestigkeit {Grundlagen} für {Ingenieure}, 3rd
  Edition, Springer-Verlag, 2007.

\end{thebibliography}

\end{document}